\documentclass[11pt]{amsart}

\usepackage{bm}
\usepackage{color}
\usepackage{amsmath}
\usepackage{amssymb}
\usepackage{amsxtra}
\usepackage{amscd}
\usepackage{amsthm}
\usepackage{amsfonts}
\usepackage{eucal}
\usepackage{latexsym,amsthm}
\usepackage{tikz} 
\usetikzlibrary{arrows} 

\newcommand{\nc}{\newcommand}
\nc{\on}{\operatorname}
\nc{\ol}{\overline}
\nc{\wt}{\widetilde}
\nc{\Wick}{{\mathbb :}}
\nc{\R}{{\mathbb R}}

\theoremstyle{remark}

\theoremstyle{definition}

\newcommand{\beq}{\begin{equation}}
\newcommand{\eeq}{\end{equation}}
\newcommand{\bmul}{\begin{multline}}
\newcommand{\emul}{{\end{multline}}}
\newcommand\beqa{\begin{eqnarray}}
\newcommand\eeqa{\end{eqnarray}}
\newcommand\bea{\begin{array}}
\newcommand\eea{\end{array}}
\newcommand\ba{\begin{array}}
\newcommand\ea{\end{array}}
\newcommand{\nn}{\nonumber}

\newcommand{\neqa}{\nonumber\end{eqnarray}}

\newcommand{\eq}[1]{eq.(\ref{#1})}

\newcommand{\ur}[1]{(\ref{#1})}

\renewcommand{\d}{\partial}
\renewcommand{\O}{{\mathcal O}}
\newcommand{\C}{{\mathbb C}}

\renewcommand{\L}{{\mathcal L}}

\newcommand{\<}{{\langle}}
\renewcommand{\>}{{\rangle}}

\nc{\CH}{{\mathcal H}}
\nc{\Db}{{\bar D}}
\nc\comment[1]{}

\newcommand{\cM}{{\mathcal M}}
\nc{\CM}{{\mathcal M}}
\nc{\CN}{{\mathcal N}}

\newcommand{\re}{\relax{\rm I\kern-.18em R}}

\renewcommand{\)}{\right)}
\renewcommand{\(}{\left(}
\nc{\al}{{\alpha}}

\def\ev{{\mathrm ev}}
\def\eps{{\epsilon}}
\def\cO{{\mathcal O}}
\def\CO{{\mathcal O}}

\def\diff{{\mathrm{Diff}}}
\nc\red[1]{{ #1}}
\nc\green[1]{}

\begin{document}
\def\k{{\kappa}}
\title{New observables in topological instantonic field theories}

\author{Andrey Losev}

\address{Institute of Systems Research and Institute of Theoretical and Experimental Physics,
Moscow, Russia}

\author{Sergey Slizovskiy}

\address{Uppsala university,
Sweden, Petersburg Nuclear Physics Institute, Russia   and   Loughborough University, UK   }
\centerline{\hfill \ UUITP-26/09  }
\centerline{\hphantom{void}}
\centerline{\hphantom{void}}

\date{March 2010}

\begin{abstract}
Instantonic theories are quantum field theories where all correlators are determined by
integrals over the finite-dimensional space (space of generalized instantons).
We consider novel geometrical observables in instantonic topological quantum mechanics that 
are strikingly different from standard evaluation observables.
These observables allow jumps of special type of the trajectory (at the point of insertion of
such observables).
 They do not (anti)commute with evaluation observables and raise the dimension of the space
of allowed configurations, while the evaluation observables lower this dimension.
We study these observables in geometric and operator formalisms.  
Simple examples are explicitly computed; they depend on linking of the points.

The new ``arbitrary jump'' observables  may be used to construct correlation functions
computing e.g. the linking numbers 
of cycles, as we illustrate on Hopf fibration.  

We expect that such observables could be generalized in an interesting way to instantonic
topological theories in all dimensions.
\end{abstract}
\maketitle

\tableofcontents

\section{Introduction}
Instantonic field theories were introduced and studied in \cite{FLN1,FLN15,FLN2}.
These  supersymmetric theories are defined by localization on instanton  space.
 These $Q$-supersymmetric theories may be considered as an extension of Witten's
topological theories \cite{Witten}  including all local observables (not necessarily $Q$-closed).

\red{We consider the class of such theories, where $Q$ is the de Rham differential on the target manifold and fermions
are identified with differentials.
In particular, we study geometric topological Quantum Mechanics, where Hamiltonian is given by Lie derivative along the given vector field.}

In this paper we introduce
a large class of $Q$-closed local observables in topological Quantum Mechanics.

One of the possible constructions is to associate observables to fibrations of the
target space.  Another possibility is to associate observables to cycles in the group of
diffeomorphisms.
All these observables do not commute with evaluation observables.
We show that even the simplest observable of this type -- corresponding to $U(1)$
fibrations -- appears in 
natural problems of geometry.
\green{and even 2D QFT.}

We start this paper by quick reminder of the formalism of
instantonic topological theories, evaluation and vector field observables in Section 2. 
This section is borrowed from \cite{FLN15}.

Novel results start in Section 3,
where we will address the  question: 
how  to write down geometrical observables
in topological quantum mechanics that do not commute
with the evaluation observables and are non-zero in
cohomologies. One type of such observables
(corresponding to diffeomorphisms that can not be
connected to identity) is well known.
Are there any other observables that have geometrical
meaning?

One way to find appropriate generalization is to note that non-trivial diffeomorphisms
correspond just to zero cycles in the space of all diffeomorphisms,
and we may generalize this to an arbitrary cycle in the space of diffeomorphisms.

Another generalization arizes if we treat allowed diffeomorphisms as allowed
jumps of the trajectory (at the prescribed time).  Simple inspection shows that
this may be generalized to jumps along compact fibers of an arbitrary fibration. 

These two classes of generalizations have one common representative that we will
study in some detail in this paper -- it corresponds to the $U(1)$ fibration.
From the point of view of cycles in diffeomorphisms we study the $S^1$ in this space,
corresponding to the $U(1)$ action on the total space of the fibration.

We think that it is instructive to discuss such observable starting with the 
vector field that generates the $U(1)$ action.
Vector field observables are $Q$-exact and seem to be irrelevant for the purpose of
constructing
nontrivial observables since they are zero in cohomology.  
However we may use them in construction of
$\alpha$-jump operators, corresponding to the $U(1)$ rotation by the angle $\alpha$.

Still such operators are equivalent to unity. To get  the novel operators we first
supersymmetrize the space $S^1$  of angles $\al$  and construct a super-jump operator,
parameterized by such superspace.  That is, the super-jump operator turns out to be  a 
differential form on $S^1$.  It is easy to show that the integral of super-jump against 
a cycle is a $Q$ - closed operator. 
 In particular, the 0-cycle, corresponding to a point $\alpha$, gives the 
$\alpha$-jump itself. While the super-jump, 
corresponding to a fundamental cycle of the circle,
has no reason to be trivial in  cohomologies of $Q$. This operator will be denoted $K$
and is a prototype of the main object of study in this paper.

The above construction may be generalized as follows.  
Consider a finite-dimensional cycle $C$ in the group of
diffeomorphism of the target $X$.  Take the operator
that pulls back the forms on $X$ to $C \times X$ \red{along the diffeomorphism action on $X$} and then integrates over the cycle $C$.  

In Section 4 we present the simplest example of correlation functions with the  observable $K$ 
and justify our expectations that it is non-trivial in cohomologies and does not commute
with evaluation observables. It follows from non-commutativity that the
correlation
functions may depend on order of times. Hence we may get worldsheet linking numbers.

In Section 5 we discuss integrated observables (integrated against time). 
These are commonly known as ``descent observables''
\cite{WittenDescent}. 
They correspond to deformations of $Q$ and hence \red{of } the Hamiltonian. 
Geometrically they correspond to 
counting  intersections
that may happen at arbitrary time.

We  consider deformations $Q \to Q + \tau K$ and 
compare them to Novikov-Witten deformation $Q \to Q + \tau \omega$.
Note that the cohomology of the former differential are just equivariant cohomology of
 the fibration.   

We 
see that in case of Novikov-Witten deformations the higher differential
corresponds to trajectory passing successively through cycles, while
in case of deformation with $K$  the higher differential   
corresponds to trajectories with successive jumps. 

 
In Section 6 we present conclusions.

\section{Sketch of geometric formalism in quantum mechanical instantonic theories}
\subsection{Idea of geometrical formalism (zero-dimensional instantonic field theory)}
Let $X$ be a
finite-dimensional manifold, $V_X$ a vector bundle over $X$, and
$v$ a section of $V$. We will call it the defining vector field. 
Then 
\begin{equation}
\<F(x,\psi)\> = \int {\rm d}p_a {\rm d}\pi_a {\rm d}x^i {\rm d}\psi^i \
\exp \left( i p_a v^a(x) -  i \pi_a \partial_j v^a  \psi^j \right)
F(x,\psi) = \int_{{\rm zeroes\ of} \ v} \omega_F
\label{euler}
\end{equation}
where $\omega_F$ denotes the differential form on $X$ corresponding to
the function $F$ on the $\Pi TX$ (with even coordinates $x^{i}$ and
odd coordinates $\psi^{i}$).  The variables $p_a$ and $\pi_a$
correspond to the even and odd coordinates on $V$.

Let us now deform $v$. In other words, let
\begin{equation}
v_{\eps} = v_{0} + {\eps}^{\alpha} v_{\alpha},
\end{equation}
where $v_{0}$ and $v_{\al}$ are sections of $V$, and $\eps \in \C^n$ are
(formal) deformation parameters. 

Consider $X \times \C^n$, and call a projection to the first factor by $pr_X$
and call by $pr_\eps$ a projection to $\C^n$.
The 
space of zeroes of $v_\eps$ for all values of $\eps$ we call  the {\it extended
instanton space} $\cM_{ext}$. It's immersion into $X\times \C^n$ we denote by  $\iota$:   
$\cM_{ext} \hookrightarrow^{\hskip -3mm \iota} X \times \C^n$.
The space $\cM_{ext}$ is fibered over $\C^n$ with projection given by $pr_\eps \circ \iota$;
the fibers $\cM_\eps$ of this fibration are zeroes of $v_{\eps}$ for given $\eps$.

\begin{tikzpicture} 
 \path (0,2) node (cM) {$\cM_{ext}$} 
 (2,2) node (XC)  {$X \times \C^n$} 
 (2,0) node (Cn) {$\C^n$} 
 (4,2) node (X) {$X$}; 
\draw[->] (XC)--node[above] {$pr_X$}(X); 
\draw[->] (XC)--node[right]{$pr_\eps$}(Cn); 
\draw[right hook->] (cM)--node[above]{$\iota$}(XC); 
\draw[->] (cM)--node[above,sloped]{$pr_\eps \circ \iota$}(Cn); 
\end{tikzpicture} 

Given a form $\omega_F$  on $X$ we may consider it as form on $X \times \C^n$ 
(it is  just $pr_X^* \omega_F$).  
Now 
we restrict it to $\cM_{ext}$ (so, we get $ \iota^* pr_X^* \omega_F $), and integrate the
resulting form against the
fibers $\cM_\eps$ of projection $pr_\eps \circ \iota$  (the operation of direct image $(pr_\eps \circ
\iota)_*$).
 This way we get a form on the base $\C^n$.  
The whole operation corresponds to multiplying $\omega_F$ (considered as a form on $X
\times \C^n$)  by the $\delta$ form on $X \times \C^n$ that
localizes to zeroes of $v_\eps$  and integrating the result over the fiber.
The integral representation of $\delta$-form on $X \times \C^n$ is built exactly as in
\eq{euler} (we simply replace $X$ with $X \times \C^n$ there):
\begin{equation}
  \int {\rm d}p_a {\rm d}\pi_a {\rm d}x^i {\rm d}\psi^i \
\exp \left( i p_a v_\eps^a(x) -  i \pi_a \partial_j v_\eps^a  \psi^j  - i \pi_a d \eps^{\alpha}
v^a_\al  \right)
F(x,\psi) =  \int_{\cM_\eps} \omega_F \equiv \hat \omega_F
\label{eulerext}
\end{equation}
In a more rigorous language
\beq
\hat \omega_F \equiv \int_{\cM_\eps} \omega_F = (pr_\eps \circ \iota)_* \iota^* pr_X^* \omega_F
\eeq

Acting with
Lie derivative $\L_\frac{\d}{\d \eps^\al} $  we get
$\CO_{v_\al}$ observable, defined as 
\begin{equation} \label{def Ov}
{\CO}_{v_{\al}} = i p_a v_{\al}^{a}(x) - i \pi_a \partial_j v_{\al}^{a}  \psi^j
 \end{equation}
 or  acting  with substitution  $ \iota_\frac{\d}{\d \eps^\al} $ we get $\pi_{v_\al}$ observable:
 \begin{equation} \label{def piv}
\pi_{v_{\al}} = i\,  \pi_a v_{\al}^{a}
 \end{equation}
    
So $\hat \omega_F$ is a generating function
for $\pi_{v_{\al}}$ and ${\CO}_{v_{\al}}$ observables
\footnote{For definition of these observables it is sufficient to consider the 
formal neighborhood of zero in $\C^n$ rather than the full $\C^n$.}
:
\beq
\< \pi_{v_{\al_1}} ... \pi_{v_{\al_l}}  \CO_{v_{\al_{l+1}}} ... \CO_{v_{\al_{n-l}}} F(x, \psi)\> = \left.  
\iota_\frac{\d}{\d \eps^{\al_1}} ... \iota_\frac{\d}{\d \eps^{\al_l}}  \L_\frac{\d}{\d \eps^{\al_{l+1}}} ...
\L_\frac{\d}{\d \eps^{\al_{n-l}}} \hat\omega_F \right |_{\eps=0} 
\eeq

The main idea of the geometrical definition of correlators in infinite dimensional case is
to consider infinite dimensional version of the above statements as the {\it  definition}
of the generating function for the correlators.

\subsection{Three points of view on instantonic quantum mechanics}
\subsubsection{Geometrical formulation of instantonic QM}
For geometrical definition of correlation function we need the following 
data: the space $X$, the differential form
$\omega$ and the defining vector field together with it's $\eps$ deformations.
Quantum mechanics is a one-dimensional quantum field theory, so we 
consider a vector field on the space of parametrized paths $\gamma$ in the target space,
\beq
\gamma \in Maps([0,T], X)
\eeq
with appropriate boundary conditions; say, $\gamma(T) = \gamma(0)$ for periodic maps 
or $\gamma(0) \in C_{in}$ and $\gamma(T) \in C_{out}$ (where $C_{in/out}$ are cycles in $X$).

The defining vector field $V_0$ gives a set of equations 
describing the evolution along the vector field $V_0$ on $X$: 
\beq \label{def equation} 
dX^i = dt \, V_{0}^{i}(X(t)) 
\eeq

Local observables  come from the evaluation map, namely, 
\beq
\ev_t : \gamma \mapsto \gamma(t)
\eeq
So for any differential form $\omega$ on the target space we may consider its
pullback to the space of parametrized paths, that we denote as $\omega(t)$:
\beq
\omega(t)=\ev_{t}^{*} \omega
\eeq 
and a general evaluation observable, corresponding to $\omega_F$ above,  is a product
of local evaluation observables at various times $\omega_F = \ev^*_{t_1} \omega_1 ... \ev^*_{t_m}
\omega_m$.  

To define a deformation  \ur{def equation} we pick up vector fields $v_\al$ and 
put them at times $t_\al$ as 
\beq \label{def equation2}
 dX^i = dt \, V_{0}^{i}(X(t)) 
+ \sum_\al \eps_\al \,  \delta(t-t_\al) \, v_\al^i(X(t))
\eeq
then we may introduce local observables ${\CO}_{v}$  and $\pi_v$.
Note, that geometrically the deformation \ur{def equation2} corresponds to jump
 of the trajectory at $t=t_\al$ by diffeomorphism that is the flow along the vector field
$v$ during the time $\eps_\al$, i.e. to 
$e^{\eps_\al {\L}_{v_\al}}$, where ${\L}$ is the Lie derivative on $X$.

We would like to stress that this aready defines basic set of correlators 
$$
\<\pi_{v_1}(t_1) ...\pi_{v_m}(t_m) \cO_{v_{m+1}}(t_{m+1}) ...\cO_{v_k}(t_k)   \omega_F \>
$$
 in the theory in finite-dimensional terms.

We may define more local observables by fusing the generating ones, namely,
given two local observables ${\CO}_{1}$ and ${\CO}_{2}$ we may define
correlator of ${\CO} _{1*2}(t_1)$ as follows:
\beq
\<{\CO}_{1*2}(t_1) \ldots \> =\lim_{t_2 \rightarrow t_1+0} 
\<{\CO}_{1} (t_1)  {\CO}_{1} (t_2) \ldots \>
\eeq

\subsubsection{Functional integral representation}
The langauge of functional integrals in a fashionable way to represent QFT, despite
it is rarely rigoriously defined. Therefore it is instructive to represent instantonic QM 
in this language. This representation is just 
the l.h.s. of \eq{eulerext}, symbolically
\beqa
& \int DX^i(t) D\psi^i(t) Dp_i(t) D\pi_i (t) \times\nn  \\
& \times e^{\int i p_j \(\frac{d X^j}{d t} -V_{0}^{j}\) - i \pi_j \(\frac{d X^j}{d t}-\partial_k V_{0}^{j} \psi^k \) dt}
F_1(X,p,\psi , \pi)(t_1) ...F_n(X,p,\psi , \pi)(t_n) \nn
\eeqa
The measure is considered to be Beresin canonical supermeasure, one may hope that
 due to balance between bosons and fermions 
it is independent of the coordinate system taken.  

In naive functional integral paradigm one should consider as local observables
functions $F$ of $X$, $\psi$, $\pi$ and $p$.
However, such functions do not give well-defined observables due to noncommutativy
between $X$ and $p$ and non-anticommutativity between $\psi$ and $\pi$.

Fixing their order means that we  have to construct
 these observables by fusing generating ones.
And generating ones do have interpretation in geometric terms:
\beqa \label{fields1}
X^i(t)=\ev_t^* X^i \\
\psi^i(t)=\ev_t^* dX^i \\
i \, p_i(t)= {\CO}_{\d/\d X^i}(t) \\
i \,\pi_i(t)=\pi_{\d/\d X^i}(t)
\label{fields4} 
\eeqa

The supersymmetry generator $Q= d_X = p_i \psi^i$ is a de Rham differential, it acts as: $Q X^i =
\psi^i$
and $Q \pi_i = p_i$.
\subsubsection{Operator approach}
The operator approach to quantum mechanics has historically been the first one \cite{Dirac}.
In this approach we have a space of states ${\mathcal H}$, Hamiltonian $H$ and a set of local operators
$\Phi_i$.
The correlators are given by 
\beq
\<\Psi_{out}|\Phi_n(t_n)... \Phi_1(t_1) |\Psi_{in} \> = 
\<\Psi_{out}| e^{-(T-t_n) H} \Phi_n ...  e^{-(t_2-t_1) H} \Phi_1 e^{-t_1 H} |\Psi_{in}\>
\eeq  
where $|\Psi_{in}\> \in {\mathcal H}$, $\<\Psi_{out}| \in {\mathcal H}^*$,  $\Phi_i, H \in {\rm End}({\mathcal H})$.
In the physics of real world the space ${\mathcal H}$ is Hermitian and $H = i H_{phys}$, where
$H_{phys}$ is Hermitian.  However in the context of general one-dimensional QFT this condition
may be omitted, for example, in statistical mechanics and in theories with complex Lagrangians.
    
Instantonic QM in operator approach  is described as follows.
The space of states is the space of differential forms on the target
and Hamiltonian is just the Lie derivative along $V_0$, which is
$Q$-exact. 

In this correspondence the evaluation operators correspond to multiplication by differential
forms
while  vector field operators $\O_v$ and $i \pi_v$ correspond to the Lie derivative and 
 to operation of contraction with the vector field respectively (hence $\{d_X, \iota_v\} = \cO_v$
is a Cartan formula). 
 All operators we consider below have geometric meaning and correlation functions are 
solutions of particular geometric 
 problems. 
 
To relate the two approaches it is convenient to introduce a geometric basis on the space of
wave-forms. 
Consider a chain $C$  on the target $X$. Then we may write a corresponding $\delta$-form
localized on this chain: $\delta_C$, roughly speaking
this is a $\delta$-form in the directions orthogonal to the cycle \cite{BottTu} 
\footnote{For example, on 2D plane $(x,y)$  a form $\delta(x) (\theta(y) - \theta(y-1)) dx$
corresponds to an interval $[(0,0), (0,1)]$.}. 
The degree of this form is $\deg \delta_C = \dim X - \dim C$.  There is a property: $d_X
\delta_C = \delta_{\d C}$.  
Cycles (i.e. chains without boundaries)  correspond to closed forms and non-contractible cycles
correspond to de Rham cohomologies of $X$.  
Taking $|\delta_C\>$ as ket- vectors we can define bra- vector
as a chain itself,
then the pairing is an intersection number: 
\beq
 \<C_1 | \delta_{C_2} \> = \int_{C_1} \delta_{C_2}  = intersection(C_1,C_2)
 \eeq
 Therefore, if we compute correlator in the operator approach one may show that
in general the position operator approach coincides with the geometrical one. For example,
the evolution operator in the operator approach means that we take the incoming chain,
deform it along the flow of the vector field (corresponding to Hamiltonian) and then intersect it with the
outgoing chain. Thus, we compute the number of intersection points. However, if
we consider the set of all preimages of these intersection points under the flow we
restore the set of trajectories, starting on an incoming chain and ending on the
outgoing one, as it should be in the geometrical approach.

This geometrical approach in a form described above  is a bit naive, since intersection of chains is defined only if they are
transversal to each other. This problem may be solved by 
"smoothening" of the incoming and outgoing chains, in particular, by replacing 
chains by smooth differential forms. Chains may be considered as limits
of smooth differential forms (and intersection is computed by the integral of the wedge product). Therefore correlator in operator approach (with states
given by smooth forms) always exist, and if the chain limit may be taken, it equals to the 
correlator in geometrical approach.

If we compactify time to a circle, then in the geometric approach we compute the number of periodic trajectories (subject to some additional requirements determined by observables).
In order to compare it to the operator approach we have to cut the circle
at some moment of time to get an interval, compute an operator on the space of differential forms, 
corresponding to this interval, and
take a supertrace of it (by supertrace we mean weighting contributions of odd forms with the
minus sign).
          
\subsection{Searching for novel local geometric observables}

As it is clear from the definition of evaluation observables, they form supercommutative 
algebra,  that goes down to supercommutative algebra on cohomology.

Observables that correspond to vector fields are either not closed or exact, so
they seem to produce nothing on the level of cohomology.

Looking at these observables one may even mistakenly conclude that all geometrical observables 
form a supercommutative structure.

However, it is well-known that diffeomorphisms that cannot be deformed to identity
provide an example of nonsupercommutative geometrical observable. We see that diffeomorphisms not
connected to identity are not small deformations and in general do not allow an expansion in powers of
small deformation parameters -- thus, there is no simple expression for such observables in terms of the 
``fields'' \ur{fields1}-\ur{fields4}.  But diffeomorphisms  have a clear geometrical meaning and one can normally work with 
such observables in geometric formalism.    

Below we will generalize this example. We will find many observables 
 that  have geometrical meaning, do not commute with evaluation observables,
and decrease the degree of \red{the wave-}form.  Thus, we study here a non-perturbative completion of evaluation and 
small-deformation observables, studied in \cite{FLN1}-\cite{FLN2}.

 \section{Integrated super-jump operator and its generalizations}
\subsection{Super-jump operator} \label{circle example}
In this section we construct a new observable in operator formulation and then explain its
geometrical meaning.

Consider a jump operator, associated with a vector field  $v$ on $X$:
\beq
{\rm Jump}_{\epsilon v}= e^{\epsilon \L_v}
\eeq
Since $\L_v$ is $\{ d_X ,  \iota_v \}$,  
\beq \label{trivial jump}
{\rm Jump}-1=\{d_X , ...\}
\eeq
so we are not getting anything interesting.

In order to get something interesting we need to consider
a super-jump operator
\beq \label{SJump}
{\rm SJump}_v(\epsilon) = e^{\epsilon \L_v + d \epsilon \,  \iota_v}
\eeq
that is a differential form on the space of parameters $\epsilon$.

Note, that this operator is $d_X+d_{\epsilon}$ closed, therefore,
being integrated along the cycle in the $\epsilon$-space it gives
the $d_X$-closed operator (we remind that  $Q=d_X$ ).

We may interpret ${\rm Jump}_{\epsilon v}$ for different $\epsilon$ as integrals of the
super-jump operator against points (0 - cycles)  in $\epsilon$-space,
 corresponding to
different values of $\epsilon$. 
Since $\epsilon$ space is connected,  all of them are equivalent to zero jump, which also follows
from 
\eq{trivial jump}.

Now it is clear how to get something more interesting - we just
need to have the space of parameters with more nontrivial cycles.

The simplest choice is to consider the $\epsilon$-space being a  circle. 
It means that the action of the vector field  is lifted to the action
of the circle, i.e. it has periodical trajectory with equal
periods (that we may take to be $1$), in other terms
\beq
{\rm Jump}_v= e^{\L_v}=1
\eeq
In this case the $\epsilon$-space has a nontrivial cycle -
fundamental cycle, and we have a new operator $K_v$ defined as  integral of the
super-jump operator along this cycle
\beq \label{Kv}
K_v = \int_{\epsilon \in S^1} {\rm SJump}_v(\epsilon) =\int_{S^1} d \epsilon \, e^{\epsilon \L_v } \
\iota_v
\eeq
The geometrical meaning of insertion of $K_v$ at time $t_K$ is to allow trajectories that 
are the trajectories of the vector field $V_0$
everywhere outside $t_K$ (solving \eq{def equation}) but they may have a jump at time $t_K$ 
along the orbit of the circle action.

Later  we will see that operator $K$ is nontrivial in cohomology and does not supercommute with
the evaluation observables. However, formulas above show that it is built out of 
observables associated to vector field - how could this happen? The tricky point is that
the operator $K$ is Q-closed in a nontrivial way. It is built using non-closed operator
$\pi$, and the integrand in \ur{Kv} is non-closed.  However, the integral is closed since the vector field $v$
produces a circle action. 

Let us make simple operator computations for the case where the target space is a circle itself,
and $X$ is an angle on that circle.
Then $K \equiv K_{\frac{\d}{\d X}}$ operator acting on degree 0 forms gives zero,  and acting
on degree 1 form  gives a number, which is an integral of this form over the circle.
Now it is clear that $K$ acts non-trivially in cohomologies since it gives $1$ when acting on
delta-form $\delta(X) \psi$ (which can be non-trivial in
cohomologies of $X$), but it gives zero if it acts on the vacuum $\bm 1$ prior to $\delta(X)
\psi$.

Thus we see that $K$ is  $Q=d_X$ closed but not exact.  
In Section 4 we will use this for operator computations of correlation functions.

\subsection{Generalization 1: Projection operator}
The above construction implies the following generalization. 
Consider a projection from the target $X$ to base manifold $B$: 
\beq
pr: X \rightarrow B
\eeq
This defines a fibration and we assume that fibers are compact.

Define the   operator
$K_{fib}$ that acts on differential forms as follows:  first integrate the differential form against
fibers of $pr$ 
to get a form on $B$. Such operation is called $pr_*$ (the differentials transverse to fibers are
identified with base
differentials).  Then take  a pullback of the integrated form from $B$  back to $X$ (this we
denote by $pr^*$), thus
\beq \label{fibration def}
K_{fib} \; \omega =  pr^* pr_* \omega = pr^* \int_{fiber} \omega
\eeq 
Such an operation (anti)commutes with de Rham differential $d_X$ since both operations
$pr^*$ and $pr_*$ 
(anti)commute with $d_X$ for compact fibers without boundary, thus it acts in cohomologies.

In quantum mechanics the evaluation observables correspond to multiplication 
of the wave-function by some form (consider, e.g. a $\delta$-form), which obviously does not
commute with integration of the wave-function over the fiber.

\subsubsection*{Geometrical meaning} 
In  geometric formalism the insertion of $K_{fib}(t)$ has an effect of jump in the instanton
solution at instant $t$ to any point on  the  fiber, containing the point
$X(t)$.  So, it is an {\it arbitrary jump } along the fiber. This definition tells what is the resulting
instanton  space (space of trajectories in case of QM). 
Since all correlation functions are computed as integrals over instanton space, the definition is
constructive. 
 
\subsection{Generalization 2: Compact cycles in the group of  diffeomorphisms of $X$ }
\label{sec:diffeo}
The  example with the circle, described in Section \ref{circle example}  can be interpreted in terms
of yet another construction.  
We may consider rotations along the arbitrary angle as
a special 1-dimensional cycle  in the group of diffeomorphysms of $X$.
It turns out that the construction above may be generalized to an arbitrary cycle in this group.
 
Indeed,  consider the group $\diff X$ of diffeomorphisms of $X$, 
denote it's action on $X$ by ${ \rm Act}: \  (\diff X) \times  X \to X $.
Choose a finite-dimensional compact cycle in diffeomorphisms: $C \subset \diff X$.        

Forms on $X$ may be pulled back to $\diff X \times X$ and integrated against  the cycle $C$. 
We may define the corresponding operator
\beq
K_{C} \; \omega =  \int_C {\rm Act}^* \omega 
\eeq
In geometric formalism this construction corresponds to allowing such jumps that 
start- and end-points of the jump may be connected by a diffeomorphism in $C$.
It is clear that the action of $K_C$ in $d_X$-cohomology is independent on the continuous
deformations of $C$.

The simplest example of this construction is a point (i.e. zero-cycle) in the space of $\diff X$.
This means
that some fixed diffeomorphism is inserted. Such constructions were already studied in the
literature under the name {\it character-valued index} \cite{WittenCharacter, Goodman}. The
particular case of it for de Rham complex is known as {\it Lefschetz number}. 
 Our jump constructions reduce then 
to twisting of the boundary conditions on the world-sheet circle used in these works.

\subsection{Digression: Cutting operator}
Note that  local observable  in Hamiltonian language is an operator $V \to V$ where $V$ is a
vector space ($V = \Omega^{\bullet}(X)$ in our case).
Any operator
can be formally represented as an infinite sum of it's matrix elements: $O = \sum C_{ij}  |\psi_i
\> \<\psi_i |$. 

Now observe that a simplest  operator $K$ on $X= S^1$ can be represented as
\beq
 K = |\delta_X\> \<X| 
\eeq
where $|\delta_X\>$ corresponds to unit wave-function. 
This formula holds for arbitrary target if $K$ allows  
jumps to any point
of the target.  It is then natural to interpret such $K$ as cutting a time interval with free
boundary conditions for both 
ends of the cut.  
This hints another possible generalization.  Let us choose two cycles $C_1$ and $C_2$ on $X$
and consider the 
corresponding wave-functions $\delta_{C_{1,2}}$ which are  $\delta$-forms, corresponding to
these cycles.  

Consider the operator 
\beq
 K_{C_1,C_2} = |\delta_{C_1}\> \<C_2|
\eeq 
that cuts the time interval and creates particular boundary conditions. In geometric formalism
it enforces the trajectory to pass through cycle $C_2$ and after passing it the trajectory jumps
to arbitrary point of cycle $C_1$. 
It is easy to express $K_{C_1,C_2}$  in terms of $K$ (arbitrary jump to any point of $X$) and
evaluation 
observables:
\beq
 K_{C_1,C_2} =\delta_{C_2} \, K \, \delta_{C_1} 
\eeq  
the fermion degree of $K_{C_1,C_2}$ is  $n_f( K_{C_1,C_2} ) = \deg \delta_{C_1}  + \deg
\delta_{C_2} - \dim X = \dim X - \dim C_1 - \dim C_2 $. 
This operator obviously does not commute with evaluation observables.

\section{Examples of geometrical computation of correlators with $K$}
\subsection{Correlator with one insertion of $K$}
To have a simplest example, consider a quantum mechanics on the circle and take the target
manifold to be also a circle.
Recall that $K = K_{\frac{\d}{\d X}}$ corresponds to arbitrary jump on the circle.

Take an evaluation observable corresponding to 1-form $\omega$: $ \ev^*_{t_2} \omega =
\omega(t_2) \psi(t_2)$ and compute
\beq
\< K(t_1)
\ev^*_{t_2}\omega\> =  \int_{S^1} \omega
\eeq

Let us start with
 geometrical computation of
 $\< K(t_1)
\omega(t_2) \psi(t_2) \>  $.
 Note, that the space of
allowed trajectories
is a space of constant maps -- so it equals to $S^1$
and is compact.
If $ V_0=c$  (see \ur{def equation} for definition of $V_0$)
then the space of allowed trajectories is
 $X(t)=X(t_1)+c(t-t_1)$
 and also equals to $S^1$ (being parametrized, say, by $X(t_1)$).

When we compute evaluation observable on this space we still
get $\int_{S^1} \omega $
(it is independent of $c$ as we expected, because
$\int_{S^1} \omega(X_1+c (t_2-t_1)) 
 =\int \omega$).
The example with non-zero $c$ shows that allowing
a jump is really necessary, otherwise there are no solutions.

The operator computation for the same correlator gives
$STr (K \omega)$.
Since the image of $K$ is only constants, the computation of $STr$
reduces to
multiplying 1 by $\omega$, acting with $K$  and 
projecting to constants.
 From the multiplication table (last paragraph in Section 3.1) it follows that the result is 
$\int_{S^1} \omega$.

\subsection{Example with two $K$ observables and two evaluation observables}
From the very beginning of topological theories there was a lot of confusion about
the nature of topological observables
$\int_{C_{i}} {\CO}_{i}$, associated to cycles $C_i$ on the worldsheet. The original proposal of Witten implied that correlator should be independent under deformation of cycles in the same homology class. However, it was again Witten (in the Chern-Simons theory) who gave an example of correlators that are linking numbers. The resoltion of the confusion is in different behaviour 
of correlators of integrands 
$$
\<{\CO}_{1}(x)   {\CO}_{2}(y)\>
$$
of the observables.
If this correlator is smooth when $x$ and $y$ coincide, the correlator of topological observables
really goes to homology of cycles $C_i$.
However, if it is singular, the only allowed moves of cycle $C_1$ are in the complement
to $C_2$ in the worldsheet, so we get a linking.
This goes to dimension 1 of the worldsheet as follows.
Correlator of observables associated to points (times) may be either smooth
(supercommutative when points are interchanging their position) or not. In the latter case we
have a 1-dimensional linking, that is the dependence of the correlator on the order of points
(usially linking is defined as a pairing between d-dimensional contractible cycles in 
$2d+1$ dimensional space, in our case $d=0$).

Since the operators $K$ do not commute with evaluation observables, we expect to get  invariants,
such as linking numbers, by computing the correlation
functions. Consider two 1-forms $\omega_1$ and $\omega_2$.

 From the operator approach the linking is almost obvious since
 $K^2=0$ and    $\omega_1 \omega_2=0$ by the form degree considerations.
Still, we would like to reproduce this result
in geometrical way.
Two $K$ operators geometrically split the circle 
 in two intervals, each of these intervals may be mapped to its own point on $X$ (or a
trajectory if $c \neq 0$),
 so when each of the intervals contains $\omega$,
 the answer is  $\int_{S^1} \omega_1 \int_{S^1} \omega_2$,
and is zero otherwise.
Taking $V_0$ to be non-zero does not really change the answer.
\beq \label{corelator_of_two}
\< K(t_1) K(t_2) \ev^*_{t_3}\omega_1  \ev^*_{t_4} \omega_2 \> = \int_{S^1} \omega_1 \int_{S^1}
\omega_2\, Link((t_1,t_2), (t_3,t_4))
\eeq
where to define $Link((t_1,t_2), (t_3,t_4))$ we fix an oriented  paths  connecting $(t_1,t_2)$ 
on $S^1$ and count intersections of it with points  $t_4$  and  $ - t_3$  with signs, determined by
the relative orientation.  This gives the linking number.

\section{Integrated observables} \label{sec:descent}
We considered above the observables that were placed at fixed times, so they corresponded
to some geometrical event (like jump of prescribed type  or passing through the chain
of prescribed type)  that happened at this particular moment.
 But there is an important class of problems 
where one is interested in geometrical event  that happens at
some (unspecified) time. 
To deal with such problems we integrate over  time of insertion of observables, and these observables are called   integrated observables. 

It is instructive to compare the operator and geometric approaches to construction of such observables.

Let us begin with the most known example of  integrated evaluation observable.
Consider the evaluation observable $\ev^* \omega$, which is a form on both the time
of evaluation and the instanton space: 
$\Omega(\R_t \times \cM)$.
Explicitly, having a space of instanton solutions $X(t,m)$, the evaluation observable equals to:
$$
\ev^* \omega(X,dX) = \omega\(X(t,m),\frac{\partial X(t,m)}{\partial t} dt + 
\frac{\partial X(t,m)}{\partial m_a} dm_a\).
$$ 
where $m_a$ stand for coordinates on the moduli space.
The component  of  $\ev^* \omega$ that has zero degree along the space of times of evaluation is the fixed time
evaluation observable $\ev^*_t \omega$ 
discussed above. 
The component containing  $dt$ is the observable 
$$\ev^*_t (\iota_{\frac{\d X}{\d t}} \omega ) dt =
\ev^*_t (\iota_{V} \omega)  dt $$
 and is known as a descent observable. This observable may be integrated 
against a subspace in the space of times.
If there are no other observables  (or if correlator is smooth in the sence described above)
and if the space of times is a circle, one can integrate this observable against this circle
(that is how we get integrated observables). However, if the space of times is an interval
(or there are other observables such that the correlator is singular and we may integrate only along 
the interval of continuity) we meet the phenomena of boundary in the
space of integration and  it makes the meaning of integrated correlators more interesting -- 
they correspond to deformations of the $Q$-operator.

In order to see this we consider the operator approach in general topological quantum mechanics.

   \subsection{Integrated observables and deformations of $Q$-operator}

\subsubsection{Topological quantum mechanics as a particular case of
general topological quantum field theory}
Consider a general topological field theory. In Atiyah formulation we should consider 
manifolds with boundary. Components of boundary are labeled as incoming 
and outgoing,  and each component (incoming or outgoing)  is associated to vector space
$V_i^{in}$ or  $V_i^{out}$  respectively. 
The main object in Atiyah formulation of TFT  is  a map that associates to any manifold 
with boundary  a linear map     $I$                                                                 
\begin{equation} \label{OP}
I  \in V_{ 1}^{in} \otimes  \ldots  \otimes V_{p}^{in}   \rightarrow 
V_{1}^{out} \otimes \ldots  \otimes V_{q}^{out}
\end{equation}
that factorizes under cutting manifold into pieces.
Applying this formulation to quantum mechanics we consider intervals and associate the same vector spaces  $V$ to both incoming and outgoing boundaries.
According to Atiyah we should associate to an interval a linear operator
$$
U \in End(V)
$$ 
such that 
$$U^2=U,$$
i.e. $I$ is a projector onto some space $V_0$;
since correlators of all operators $\Phi$ are given by their restriction to $V_0$: 
 $I\Phi I$,  we may start with $V=V_0$.
This is nice but it is not exactly what we have in geometrical theories.

To include such theories in the formalism we need to extend Atiyah's formulation to
Segal's one - namely,  we have to replace manifolds by manifolds equipped with local geometrical data. 
By local data we mean the data on $X$ that uniquely determines the data on any piece 
of $X$, i.e. there is a map
$$
Cut_i :  Geom(X )\rightarrow Geom(X_i) 
$$
  As an example of such data we may take metric or complex structure.

According to Segal, the main object is a map $I$
from $Geom (X)$ to the  space  (\ref{OP}),
i.e. 
\begin{equation} \label{OP1}
I  \in V_{ 1}^{in} \otimes  \ldots  \otimes V_{p}^{in}   \otimes
V_{1}^{out,*} \otimes \ldots  \otimes V_{q}^{out, *} \otimes Funct(Geom(X))
\end{equation}
such that \red{for $X = X_1 \cup X_2$ }
\green{\begin{equation} \label{Funct}
I(X)=Cut_1^{*} I(X_1) \cdot  Cut_2^{*} I(X_2) 
\end{equation}
}
\red{\begin{equation} \label{Funct}
I(X)=Cut_1^{*} I(X) \cdot  Cut_2^{*} I(X) 
\end{equation}
}

here $\cdot$ stands for the natural contraction between vector spaces corresponding to boundaries that appear in cutting.
In application to quantum mechanics (where we take the metric on time as a local geometrical data) it means that
$$
U(t_1) U(t_2)=U(t_1+t_2)
$$
where $t_i$ are  lengths of the intervals.
This equation is  solved by
$$
U(t)=\exp (-tH),
$$
that is  a well-known evolution process in operator formulation of quantum mechanics,
where $H \in End(V)$ is a Hamiltonian (in Euclidean signature).

In order to define topological theory we replace  spaces $V$ and $Funct(Geom(X))$ by complexes.
For the space $V$ we may take the same space but with a differential  
$Q$ that squares to zero, while $Funct(X)$ has to be replaced by the space $\Omega(X)$ of
all differential forms on the space of geometrical data, so that operator $d_{Geom}$ acts on it.
The main condition for $I$ is the 
closeness of $I$ with respect to the total action:
\begin{equation} \label{topol}
(Q+d_{Geom}) I=0
\end{equation}
together with factorization condition that looks exactly like (\ref{Funct}) with space of functions being replaced by the space of differential forms on the geometrical data.

The universal solution to equation (\ref{topol})
in the case of quantum mechanics is given by
 \begin{equation} \label{topsol}
U(t,dt)= \exp (-[Q+d_t, \, t G ]) =\exp(-t H-dt G ), \ \text{ with }   H=\{  Q,  G \}
\end{equation}

Now we may define observables, we will do it here
for the case of manifold $X$ equipped with the Riemann metric.
People use to study local observables, however, we will
define here the notion of { \it subspace observables } as follows.
Consider the subspace $Y$ of the worldsheet space $X$,  and consider the $\epsilon$  tubular 
neighborhood  of $Y$ ,
 $$
Y_{\epsilon}=\{ x \in X,  dist(x,Y)<\epsilon \}
$$
where $dist (x,Y)$ is a distance between the point $x$ and the subspace $Y$,
and we will take $\epsilon$ to be small enough.

Consider $I(X \backslash  Y_{\epsilon})$, it has additional boundary formed by points 
$$
\Gamma(Y_{\epsilon})= \{ x \in X,  dist(x,Y)=\epsilon \}.
$$
This boundary contains one component when dimension of $X$ is bigger than 1, while it contains 
two components for 1-dimensional $X$. In the former case we will take the boundary to be incoming,
while in the latter case we take one component to be incoming and the second -- outgoing.
Finally, let us take the state $v_{\epsilon}$ in the multidimensional case and 
the operator $\Phi_{\epsilon}$ in the one-dimensional case such that
the $\epsilon \rightarrow 0$ limit of the contraction between $I$ and $v$ exists. So we define 
in the multidimensional case

\begin{equation}
I(X,  O(Y)_{v}) =  \lim_{ \epsilon \rightarrow 0} I(X \backslash  Y_{\epsilon}) v_{\epsilon}
\end{equation}
and in the one-dimensional case
\begin{equation}
I(X,  O(P)_{\Phi}) =  \lim_{ \epsilon \rightarrow 0} I(X \backslash  P_{\epsilon}) \cdot  \Phi_{\epsilon}
\end{equation}
where $P$ is a point and $\cdot$  stands for the contraction between the operator and  $V\times V^*$ associated to
the two boundaries of the tubular neighbourhood of the point $P$.

While peculiarities of the limit are rather interesting in the multidimensional case,
in the one-dimensional case the situation is rather simple. Therefore, the generic correlator in quantum mechanics is given by a well-known formula
\begin{equation}
\<out | U( T-t_n )\Phi_n  \ldots  U (t_2-t_1) \Phi_1 U(t_1) |in\>
\end{equation}
and the only difference in the topological quantum mechanical case is given by replacement of evolution operators 
$U(t)$ by their superanalogues.

This means that the generic correlator of local observables in quantum mechanics 
and the universal corrrelator on an interval  equals to
\begin{equation} \label{cortop}
I=\<out | U( T-t_n, dT-dt_n )\Phi_n  \ldots  U (t_2-t_1, dt_2-dt_1) \Phi_1 U(t_1,dt_1)|in\>
\end{equation}
here $t_i$ are the positions of marked points on the interval of length $T$.
One may show that 
\begin{equation} \label{dtop}
dI=0
\end{equation}
for $Q$-closed operators and initial and final states.
In particular, the zero form component is independent of $t$ -- that is the topologicity in strict sense.
The topologicity for higher forms is not that obvious -- it only means that integrals of $I$
along cycles do not depend on smooth deformations of these cycles.

\subsection{Integrated observable}

\subsubsection{Integrated descent observable and non-Q-closeness
of its integral }
Now we are in position to give the universal definition of the integrated observable -
(recall that in geomerical incarnation it stated that corresponding  geometrical event happens at nonspecified moment)
it means that we intergate the differential form (\ref{cortop}) along the position of the marked
point.  Therefore,  from the perspective of original quantum mechanics
it corresponds to insertion of the operator 
$$
\Phi_{i}^{(1)}=\{ G, \Phi_i \}
$$
 at point $t_i$ and integration of it along the time manifold.
Symbolically,  we may say that we study
$$
\< \int_X \Phi_{i}^{(1)}   \Phi_1(t_1) ...  \Phi_{i-1}(t_{i-1}) \Phi_{i+1}(t_{i+1}) ... \Phi_n(t_n)\>
$$  

Such operator was introduced by Witten as descendant operator,
since it obviously solves the descent equation
\begin{equation} \label{des}
\{ Q,  \Phi^{(1)} \} =  [ H , \Phi ] = \frac{d}{dt} \Phi
\end{equation}
where the last equality holds under correlator.

Naively, one may think that such observables preserve $Q$ -- the naive argument goes as follows:
Take $Q$-exact operator $\Phi_1= [ Q, \Psi ] $ and put it under correlator.
Take $Q$ from $\Psi$ and act with it on $\Phi_i$ -- it would give a total derivative.
Suppose that we integrate along a compact time manifold without boundaries -- then the integral
of total derivative is zero.

Naiveness of this argument shows up already when we consider time manifold
with boundaries --  in this case total derivative results in action of
operator $\Phi$ on boundary states. It makes us think that decoupling of $Q$-closed observable
happens under additional condition that boundary states are annihilated by $\Phi_i$.
Moreover, close inspection of the region of integration reveals another type
of boundaries -- when integrated operator hits  operators, placed at fixed moments 
$t_1, \ldots t_n$.  In this case the boundary contributions are expressed
as commutators
$$
[ \Phi_i, \Phi_j]
$$

\subsubsection{Homological meaning of integrated observable}

One may think that boundary contributions for integrated observables obstruct 
the homological interpretation of integrated observable. However, situation is  simpler than one may expect: 
integrated observables correspond to deformations of $Q$-symmetry.
In particular, consider deformation of $Q$ symmetry of the following form:
$$
Q_{\tau}= Q+\tau \Phi,
$$
where we assume that 
$$
\Phi^2=0.
$$
If we keep the superpartner of the Hamiltonian -- $G$ --  intact we conclude that
the zero degree component of the evolution operator changes as follows
$$
 \exp (-t (H+\tau \{ G, \Phi \} ))=
\exp(-tH)+\tau \int dt_1 \exp(-(t-t_1) H )  \{ G, \Phi \} \exp(-t_1 H) + \ldots
$$
and 1-form component is not changing (here we also assume that $G^2=0$)
i.e. we just have the generating function for integrated observable with generating parameter
$\tau$. 
 Now we may easily interpret the boundary contributions -- they correspond to the action
of  $Q_{\tau}$ on states and observables, and vanishing of boundary terms means that such states and
observables are annihilated by the family of operators $\Phi$.

But this is not natural -- rather one would expect that there is a family of operators
$\Phi_{\tau}$  or a
family of states  annihilated by $Q_{\tau}$:
\begin{equation} \label{def}
(Q+\tau \Phi ) ( |in_0\>+ \tau | in_1\>  +\tau^2 |in_2\> + \ldots )=0
\end{equation}
It is easy to show that taking into account the change of initial state $|in_1\>$
we cancel the non-closeness of the integrated observable.

However, even this is not the end of the story -- below we will show 
that there are obstructions in finding of such families.
Moreover, these obstructions are also expressed in terms of integrated correlators.

\subsubsection{Obstructions and integrated  correlators}
Consider the problem of construction of perturbative family of $Q_{\tau}$ closed
states like in (\ref{def}),
   modulus  $Q_{\tau}$ exact states. Clearly, 
$|in_0\>$ should be a representative of $Q$-cohomology class.
What about $|in_1\>$?  It should be a solution to
\begin{equation} \label{fo}
Q  |in_1\>=\Phi  |in_0\>
\end{equation}
The right hand side of (\ref{fo}) is $Q$-closed while the
equation itself states that stronger statement holds -- it is $Q$-exact.
The obstruction for this belongs in the cohomology class of the right hand side of (\ref{fo}),
i.e.  it is measured by 
\begin{equation} \label{ofo}
Obstr_1= \<out_0 | \Phi | in_0\>
\end{equation}
where $\<out_0|$ is an element of the dual space of states representing 
a generic class of $Q$-cohomology.
If the obstruction equals to zero we may proceed to the second order problem where we compute
\begin{equation} \label{so}
Q |in_2\>=\Phi |in_1\>= \Phi Q^{-1} \Phi |in_0\> 
\end{equation}
and the second order obstruction equals to
\begin{equation} \label{oso}
Obstr_2=\<out_0|  \Phi Q^{-1} \Phi |in_0\> .
\end{equation}

In the case of topological quantum mechanics there is a natural candidate for 
$Q^{-1}$, namely, let us take 
\begin{equation} \label{hodge}
h_{QM}= \int_{0}^{+ \infty} G \, dt \, e^{-t H}
\end{equation}
If $H=\{  Q , G \}$ satisfies the Hodge condition, i.e.  it is positive definite outside the cohomology
and vanishes on the cohomology, then the integral in the right hand side of (\ref{hodge}) exists
and 
\begin{equation} \label{homot}
\{ Q,  h_{QM}   \} =    1- \Pi,
\end{equation}
where $\Pi$ is the projector on the space of zero modes of $H$.
It means that
$$
Q h_{QM} \Phi |in_0\>=\Phi |in_0\>  - \Pi  \Phi |in_0\> =\Phi |in_0\>
$$
where the second equality holds  when the first obstruction vanishes,
so $h_{QM}$ really works as $Q^{-1}$.

This construction is called Hodge construction since it was extensively studied on the example of
de Rham cohomology of compact Riemann manifold. In this case 
$$
G=d^* \ \text{ and }  H=\Delta
$$ 
such topological quantum mechanics is well-known as ${\mathcal N}=1$ supersymmetric
quantum mechanics.

It could be that $h_{QM}$ may serve as $Q^{-1}$ even if Hodge condition is not satisfied.
To see this we consider
\begin{equation} \label{spec_homotopy}
 Q \int_{0}^{+ \infty} G\, dt \, e^{-t H} \Phi |in_0\>=
\Phi |in_0\>  -  e^{- \infty H} \Phi |in_0\>
\end{equation}

Therefore, in this case $h_{QM}$ may work as inverse $Q$ if
 the limiting action of   $\exp (- \infty H)$
 on the state  $\Phi |in_0\>$ does not only exist but also equals to zero. 

Interestingly enough this may happen in geometrical quantum mechanics 
where the Hamiltonian is the Lie derivative. In general, vector field may have 
limiting cycles (this may be cured by considering Morse vector field), and still 
the limiting action of the Morse flow may be non-vanishing.
However, we will encounter below the example where everything works. 

All this means that it is reasonable to consider the following correlator in topological quantum mechanics
\begin{equation} \label{hcor}
\<u_0|  \Phi  h_{QM} \Phi |v_0\>= 
 \int_{0}^{+ \infty}  \<u_0|  \Phi  G\,  dt\, e^{-t H} \Phi |v_0\>
\end{equation}
that under condition discussed above leads to the second obstruction to solution of
homological problem (\ref{def}).

From the point of view of general topological quantum mechanics it is  an 
integral over the space of metrics  on an interval.  Such object is often called an answer in topological
gravity since we integrate against the space of metrics on a space-time, that is time in our case. 
From the point of view of geometrical topological theory it means that some geometrical event
(given by  the action of $\Phi$  on $|in_0\>$)   has happened at the beginning of time,
then evolution took place until the second event happened
(given by $\<out_0| \Phi$).


\subsection{Geometrical examples of deformation of $Q$}
\subsubsection{Massey operations}
The first example of deformed operator $Q$ comes from
evaluation observables. In this case we
consider Witten-Novikov operator
$$
d +\tau \omega
$$
where differential form corresponds to evaluation observable associated to
$\omega$.

Interestingly, obstructions (starting from the second one)  that we mentioned above correspond to Massey
operations. In particularly, it means that they may be computed in
geometrical formulation of quantum mechanics,
i.e. in terms of number of trajectories of the vector field passing through
cycles (associated to differential form $\omega$).
In this sense we see that higher obstructions are nothing but one-dimensional
analogues of the celebrated Gromov-Witten invariants that compute the number of
holomorphic curves passing through the prescribed set of cycles.
We will discuss it in more details elsewhere, but it is not the
main topic in the present paper -- here we would like to concentrate on 
$K$ operators, that correspond to integrated jumps.

\subsection{Equivariant cohomologies and jump operators}
It turns out that geometrical problems associated to arbitrary jump operator $K$ arise in computation of
equivariant cohomology.

Suppose that we have a $U(1)$ bundle $X$ with the base $Y$. 
One may study equivariant cohomology, i.e. cohomology on the space of $U(1)$ invariant forms
with differential
\beq
Q_{eq}=d + \tau \iota_{v} \ ,
\eeq 
where the vector field $v$ generates the $U(1)$ action.
It is known that equivariant cohomology in the space of differential forms taking values in
polynomials in $\tau$ are related to the cohomology of the base as follows:
one has to substitute $\tau$ with the first Chern class of the bundle, i.e. with the
class of curvature of the $U(1)$ connection.

In the case of integrated $K$ observable we should study the operator 
\begin{equation} \label{newd}
Q_{def} = d + \tau K 
\end{equation}
  acting on the space of all differential forms.
Since $K$ involves integration along the fiber it projects forms to invariant ones.
It seems that people have missed the operator (\ref{newd}) since it is not differential
operator, but we pay attention to it since it is geometrical.
 
Really, computation of obstructions for such new operator turns out to be an 
interesting geometrical problem in geometrical quantum mechanics.
In particular, we may consider a Hopf bundle,  that is a sphere $S^3$ fibered over a sphere 
$S^2$. 
Let us compute the second obstruction for deformation of the three-form that is a delta-function
on a point that we will call $P$.
It is clear that the first obstruction vanishes. Really,  the action of $K$ on the three-form gives a 
2-form that is a delta function on a fiber passing through this point. Since 
all 2-cycles on a three-sphere are contractable the first obstruction vanishes.

The quantum mechanical expression for contraction provides a more detailed information
on how this contraction happens.
Really, consider as a Hamiltonian the  special vector field $V_0$ on a three-sphere that leaves one point invariant and
contracts the rest of the sphere to another point such that these fixed points of the vector fields
do not coincide with the point $P$. 
\green{One may show that} The integral
$$
\int_{0}^{T} \exp(-tH)\, G \, dt  \, \delta_{Fiber_P}
$$
is given by a 1-form delta-form on an annulus formed by evolution lines of the special vector 
field $V_0$  that happens in time $T$ and that starts on the fiber passing through the point $P$.
When $T$ goes to infinity this annulus tends to a disc (and the fiber passing through the point
$P$ is its only boundary).
It means that conditions of special homotopy (see \ur{spec_homotopy} and below) hold.

Now we need to apply $K$ operator to it and intersect with the outcoming cycle.
However, geometrically it is more convenient to apply $K$-operator to the outcoming cycle
and intersect it with the disc.

Really, if we take another point $R$  as an outcoming cycle then the action of $K$ on it
gives the delta-function on the fiber passing through the point $Q$. Therefore,
the second obstruction equals to intersection of the fiber passing through the point $Q$
and the disc, who's boundary is the fiber passing through $P$, i.e.
it equals to linking number between fibers. This number equals to 1 for Hopf fibration. 

Putting everything together, we get
\begin{equation} \label{example} 
 \int_{0}^{+ \infty}  \<R|  K \, \iota_{V_0} dt e^{-t \L_{V_0}} K |P\>=
Fiber_R \cap  Disc_P= Link (Fiber_R, Fiber_P)=1
\end{equation}

Geometrically, the only trajectory contributing to the correlator looks as follows:
it starts at point $P$, jumps along the fiber, then it moves along the trajectory of vector field
over the disc towards the intersection with the second fiber. At this point trajectory jumps
again to point $R$.
That is how jump operators reveal themself in computations in equivariant cohomology
(really, in a problem equivalent to computation of equivariant cohomology).

\subsection*{Acknowledgements}
The work of A.L. was supported by grant for support of Scientific Schools 
LSS-3036.2008.2  and RFBR grant 
07-01-00526.
S.S. acknowledges support and encouragement from Prof. Antti Niemi,  STINT Institutional Grant
and VR Grant 2006-3376.

\end{document}